\documentclass[11pt,amsmath,amssymb]{revtex4}
\pdfoutput=1

\usepackage{graphicx}
\usepackage{amsmath,amssymb}
\usepackage{color}
\definecolor{darkblue}{rgb}{0,0,0.5}
\definecolor{lila}{rgb}{0.3,0,0.3}
\definecolor{turq}{rgb}{0,0.1,0.4}

\usepackage[pdftex,
colorlinks=true,
 linkcolor=darkblue, 
 filecolor=red,
 citecolor=turq, 
 urlcolor=lila, 
 pdftitle={Exploring the limits of single emitter detection in fluorescence and extinction},
 pdfauthor={Gert Wrigge, Jaesuk Hwang, Ilja Gerhardt, Gert Zumofen, Vahid Sandoghdar},
 pdfsubject={Signal to Noise Ratio in Single Molecule Spectroscopy},
 pdfkeywords={Single Molecules, Resonance Fluorescence, Signal-to-Noise Ratios, SNR, Fluorescence Spectroscopy, Extinction Spectroscopy, Extinction, Absorption, Weakly Emitting Systems, Solid Immersion Lens, SIL, Rare Earth Ions, Homodyning},
 pdfpagelabels=true,
 breaklinks=true,
 plainpages=false,
 bookmarks, bookmarksnumbered=true
]{hyperref}

\begin{document}

\title{Exploring the limits of single emitter detection in fluorescence and extinction}

\author{G. Wrigge, J. Hwang, I. Gerhardt\footnote{Current address: \href{http://quantumlah.org}{CQT, Centre for Quantum Technologies}, 3 Science Drive 2, 117543 Singapore}, G. Zumofen, and V. Sandoghdar}
\email{vahid.sandoghdar@ethz.ch}

\affiliation{\href{http://www.nano-optics.ethz.ch}{Laboratory of Physical Chemistry} and \href{http://www.opteth.ethz.ch}{optETH}, ETH Z\"{u}rich, CH-8093 Z\"{u}rich, Switzerland}

\begin{abstract}
We present an experimental comparison and a theoretical analysis of
the signal-to-noise ratios in fluorescence and extinction
spectroscopy of a single emitter. We show that extinction
measurements can be advantageous if the emitter is weakly excited.
Furthermore, we discuss the potential of this method for the
detection and spectroscopy of weakly emitting systems such as rare
earth ions.
\end{abstract}

\maketitle

The progress of nanoscience and technology in the past two decades
has been accompanied by a growing interest in the optical study of
single nano-objects~\cite{Kulzer:04}. A major thrust in this
research area came from cryogenic
spectroscopy~\cite{Moerner:89,Orrit:90} as well as room temperature
detection~\cite{Shera:90} and microscopy~\cite{Betzig:93,Nie:94} of
dye molecules. Although a fluorescent atom suspended in vacuum can
be seen even by the naked eye, achieving a high signal-to-noise
ratio (SNR) in the detection of single molecules is a nontrivial
task in the condensed phase. In particular, the background light and
noise associated with the fluorescence or scattering from the
environment can easily dominate the small signal of a single
emitter. Furthermore, the dark counts and noise of photodetectors
put a limit on the lowest signals that one might hope to detect.

\begin{figure}[htb]
\centerline{\includegraphics[width=8.5cm]{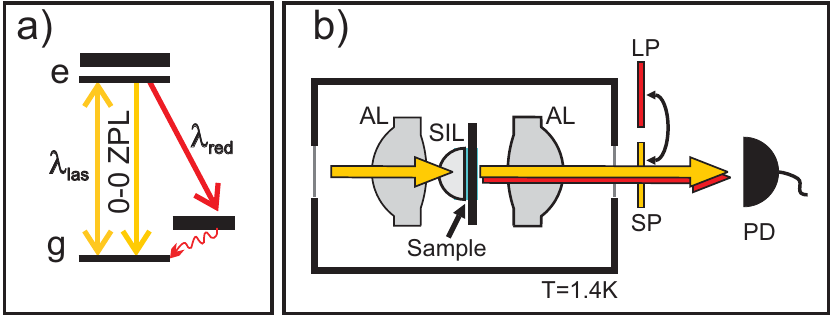}}
\caption{a) The level scheme of a dye molecule. b) The schematics of our
experimental setup. A laser beam is focused onto the sample using an
aspheric lens (AL) and a hemispherical solid-immersion lens (SIL). A
second aspherical lens is used to collect the transmitted laser beam
as well as the forward emitted fluorescence of the molecule. LP:
long pass filter, SP: short pass filter, PD: photodetector.
\label{setup}}
\end{figure}

As shown in Fig.~\ref{setup}a, the level scheme of a fluorescent
molecule consists of vibrational manifolds in the electronic ground
($g$) and excited ($e$) states. For an appropriate combination of an
emitter and its surrounding matrix, the linewidth of the so-called
zero-phonon line (ZPL) of the 0-0 transition between the vibrational
ground states of $g$ and $e$ can become lifetime limited at
cryogenic temperatures, thus enhancing the emitter's absorption
cross section $\sigma$~\cite{SMbook}. A very successful method for
detecting a single molecule with a narrow 0-0 ZPL has been
fluorescence excitation spectroscopy~\cite{Orrit:90} where the
red-shifted incoherent fluorescence of the molecule at wavelength
$\lambda_{\rm red}$ is separated from the light at the laser
wavelength $\lambda_{\rm las}$ by using high quality spectral
filters. The SNR of this technique is determined on the one hand by
the detector noise, which can be as low as 20-100 counts per second
(cps) for very good avalanche photodiode single photon counters. On
the other hand, saturation limits the maximum attainable signal to
typical values of $10^5-10^6$~cps on the detector for a good dye
molecule. Thus, fluorescence excitation spectroscopy can enjoy a
very healthy SNR when applied to strongly fluorescent systems.
Detection of very weak emitters, however, remains a challenge. In
particular, fluorescence detection of single rare earth ions has
been hampered owing to their long lifetimes and therefore ultra weak
fluorescence.

An alternative approach to the detection of single solid-state
emitters is to go back to the first method that was applied in
single molecule spectroscopy~\cite{Moerner:89}, namely to detect the
extinction of the laser light caused by a single molecule in its
path. This method was successfully revived by Plakhotnik and Palm in
2001~\cite{Plakhotnik:01} where the coherent scattering of the
excitation light was interfered with the residual reflections from
the interfaces in the setup. Closely related efforts followed on
quantum dots, especially with the aim of acquiring access to the
linewidth of the main optical transition in these
systems~\cite{Guest:02,Alen:03}. Recently, we have extended this
approach to detect the extinction of a laser beam by a single
molecule in transmission without the need for any noise suppression
technique~\cite{Gerhardt:07a,Gerhardt:07b,Wrigge:08}. In this paper,
we compare the conventional fluorescence excitation technique with
extinction measurements in terms of the SNR and discuss the
potential of the latter for detecting emitters with very weak
optical transitions.

\begin{figure}[htb]
\centerline{\includegraphics[width=9.5cm]{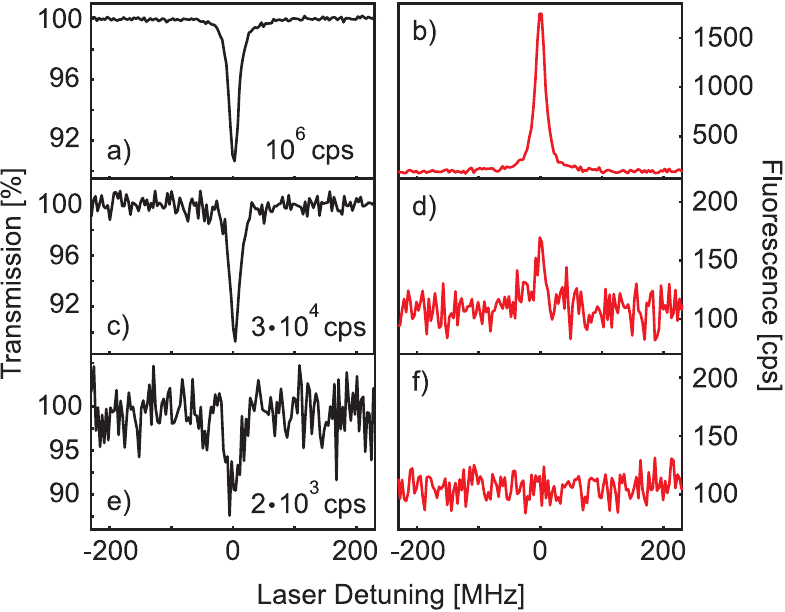}}
\caption{a), c),
e) Extinction spectra recorded from a single molecule in
transmission at three different detected laser powers of $10^6$,
$3\times10^4$, $2\times 10^3$ cps as measured on the
detector. b), d), f) Fluorescence excitation spectra recorded under
same conditions as spectra a), c), and e), respectively.
\label{spectra}}
\end{figure}

The experimental arrangement of our discussion is depicted in
Fig.~\ref{setup}b, and its details are described in
Refs.~\cite{Gerhardt:07a,Wrigge:08}. Briefly, the excitation laser
light was focused onto the sample consisting of DBATT molecules
embedded in a \emph{n}-tetradecane matrix inside a cryostat. For
this we used an aspheric lens with a numerical aperture of $0.68$
and a cubic zirconia hemispherical solid immersion lens (SIL). After
interaction with the sample, a second aspheric lens collimates the
beam and directs it to an avalanche photodiode (PD). Two different
filter sets are used to either reject $\lambda_{\rm las}$ and detect
$\lambda_{\rm red}$ or vice versa. The former arrangement delivers a
fluorescence excitation spectrum while the latter allows a direct
resonant measurement. Fig.~\ref{spectra} shows examples of
fluorescence and extinction spectra recorded from the same single
molecule and on the same detector at three different incident
powers. In this article we adopt the unit of counts per second (cps)
for power. When the detected laser power reads $10^6$ cps on PD
(corresponding to an excitation regime well below saturation) both
extinction (a) and fluorescence (b) yield comparable SNR of
$\approx$ $100~\sqrt{\rm Hz}$. For a detected laser power of $3.2
\times 10^4$ cps, the fluorescence of the molecule is hardly above
the detector dark count rate of 100 cps. However, the extinction is
still easily observable at 10\% visibility. Even at an ultra-low
illumination level of $2000$ cps the extinction signal succeeds in
detecting the molecule whereas the fluorescence peak is fully buried
under the detector noise.

Assuming a perfect transmission channel and detector, the power on
PD in the absence of any spectral filter is given in cps by
\begin{eqnarray}
\label{mainformula}
   P  &=& \frac {\epsilon_0 c}{2 \hbar \omega}
   \int_{\Omega}  \left( \left\langle  {\bf \hat E}_{\rm las}^- \cdot {\bf \hat E}_{\rm las}^+ \right\rangle
   + \left\langle  {\bf \hat E}_{\rm m}^- \cdot {\bf \hat E}_{\rm m}^+ \right\rangle
   +2 \Re \left\langle  {\bf \hat E}_{\rm las}^- \cdot {\bf \hat E}_{\rm m}^+ \right\rangle  \right)
   d\Omega  \nonumber \\
   &=&  P_{\rm las}  + P^{\Omega}_{\rm m} -  P_{\rm ext}
 \end{eqnarray}
where $E_{\rm las}$ and $E_{\rm m}$ represent the electric fields
associated with the laser and the molecular emission at the
detector, respectively, and $\omega$ is the frequency of the emitted
photon. $\Omega$ denotes the solid angle of light collection and is
assumed to cover all the transmitted laser light. The molecular
emission $P^{\Omega}_{\rm m}$ consists of a part that originates
from the 0-0 ZPL transition and is resonant with the laser light and
a red-shifted component which results from molecular and lattice
vibronic transitions. The electric field associated with the
coherent part of the resonance
fluorescence~\cite{Wrigge:08,cohen-book} gives rise to a nonzero
third term $P_{\rm ext}$ of Eq.~(\ref{mainformula}), signifying the
interference between the molecular emission and the laser beam. This
component, which is known as the ``extinction"
term~\cite{Jackson-book} is equivalent to a homodyne signal where
the excitation laser beam acts as the local
oscillator~\cite{Haus-new,Yuen:90}.

It is helpful for the following discussion to write the terms of
Eq.~(\ref{mainformula}) in an explicit manner:
\begin{eqnarray} \label{bunch}
     P_{\rm las}  &=&  \frac {\epsilon_0 c}{2 \hbar \omega}
     \int_{\Omega } \left\langle  {\bf \hat E}_{\rm las}^- \cdot {\bf \hat E}_{\rm las}^+ \right\rangle
     d\Omega \nonumber \\
     P_{\rm m}^{4\pi}  &=&  \frac {\epsilon_0 c}{2 \hbar \omega}
     \int_{4 \pi } \left\langle  {\bf \hat E}_{\rm m}^- \cdot {\bf \hat E}_{\rm m}^+ \right\rangle
     d \Omega
     =  \Gamma_1 \rho_{22} = \frac {  \Gamma_1 } 2  \frac { S} {1+S} \nonumber \\
     P^{\Omega}_{\rm m}  &=& \zeta P_{\rm m}^{4\pi} \nonumber \\
     P^{\rm res}_{\rm m}  &=&   \alpha P^{\Omega}_{\rm m}  \nonumber \\
     P^{\rm red}_{\rm m} &=&   (1-\alpha) P^{\Omega}_{\rm m} \nonumber \\
     P_{\rm ext}  &=& -\frac {\epsilon_0 c}{2 \hbar \omega}
   \int_{\Omega}  2 \Re \left\langle  {\bf \hat E}_{\rm las}^- \cdot {\bf \hat E}_{\rm m}^+ \right\rangle
   d\Omega  ~.
\end{eqnarray}
The quantity $P_{\rm m}^{4\pi}$ gives the total power emitted by the
molecule into the $4\pi$ solid angle. $\rho_{\rm 22}$ is the
population of the excited state, and the on-resonance saturation
parameter S reads~\cite{cohen-book}
 \begin{eqnarray}\label{saturation-1}
    S = \frac{{\cal V}^2 }{\Gamma_1 \Gamma_2},
\end{eqnarray}
where ${\cal V}$ is the Rabi frequency defined by $\hbar {\cal V} =
{d}_{\rm ZPL} \cdot {E}_{\rm las}({\rm O})$. The transition dipole
moment ${d}_{\rm ZPL}$ and the incident electric field ${E}_{\rm
las}(\rm{O})$ at position of the molecule are assumed to be parallel
for simplicity. The factor $\alpha$ describes the ratio of the power
emitted on the 0-0 ZPL to the total excited state emission. Thus,
${d}_{\rm ZPL}=\sqrt{\alpha}d_{\rm eg}$ where ${d}_{\rm eg}$ denotes
the dipole moment associated with the total spontaneous emission
rate of the excited state given by $\Gamma_1 = d_{\rm eg}^2
\omega^3/(3 \pi \epsilon_0 \hbar c^3)$. $\Gamma_2$ represents the
transverse decay rate which equals $\Gamma_1/2$ in the absence of
any dephasing. The parameter $\zeta$ signifies the fraction of the
total emitted molecular power to that collected into the detection
solid angle $\Omega$. We note that in addition, one might have to
account for total internal reflection and waveguiding in the
substrate which influence the angular distribution of the laser
light and the molecular emission~\cite{Gerhardt:07a}. Finally, the
quantities $P^{\rm res}_{\rm m}$ and $P^{\rm red}_{\rm m}$ represent
the portions of the molecular emission into the solid angle $\Omega$
that are resonant with the excitation laser and red shifted from it,
respectively.

It is now instructive to separate the properties of the laser beam
from the spectroscopic features of the emitter. Using the
definitions of $\Gamma_1$ and $\cal V$, one can rearrange the
saturation parameter in Eq.~(\ref{saturation-1}) to read
\begin{eqnarray}
  S = \frac \alpha {\Gamma_2} {\cal K}  P_{\rm las}
\end{eqnarray}
where ${\cal K}$ is a unitless geometrical factor that relates
$|E_{\rm las}({\rm O})|^2$ to the laser power $P_{\rm las}$. More
precisely, ${\cal K}$ denotes the ratio of the total power scattered
by a weakly excited two-level system and the incident power. It
depends on the spatial mode of the laser beam and the focusing
optics. The reader is referred to Ref.~\cite{Zumofen:08} for
details.

The expressions in Eq.~(\ref{bunch}) provide us with the red shifted
fluorescence $P^{\rm red}_{\rm m}$. The noise on this signal is
given by the fluctuations in the detector dark counts $P_{\rm drk}$
if we assume that the excitation light is completely rejected by the
filters. Thus, the SNR for a fluorescence excitation measurement
becomes
\begin{equation}
\label{SNRfluo}
 {\rm SNR}_{\rm red}  =  \frac{\mu P^{\rm red }_{\rm m} }{ N_{\rm red }  }
 = \frac{ \mu  \zeta (1-\alpha) \Gamma_1  }{2 \sqrt{  P_{\rm drk }}  }
 \frac{S}{1+S}.
\end{equation}
where we have introduced $\mu$ to account for losses (e.g. cryostat
windows, filters, etc.) and the detector efficiency. The SNR maximum
is given by $ {\rm SNR}^{\rm max}_{\rm red}=\mu \zeta (1-\alpha)
\Gamma_1 /(2 \sqrt{ P_{\rm drk }})$ and occurs in the fully
saturated regime.

\begin{figure}[htb]
\centerline{\includegraphics[width=8cm]{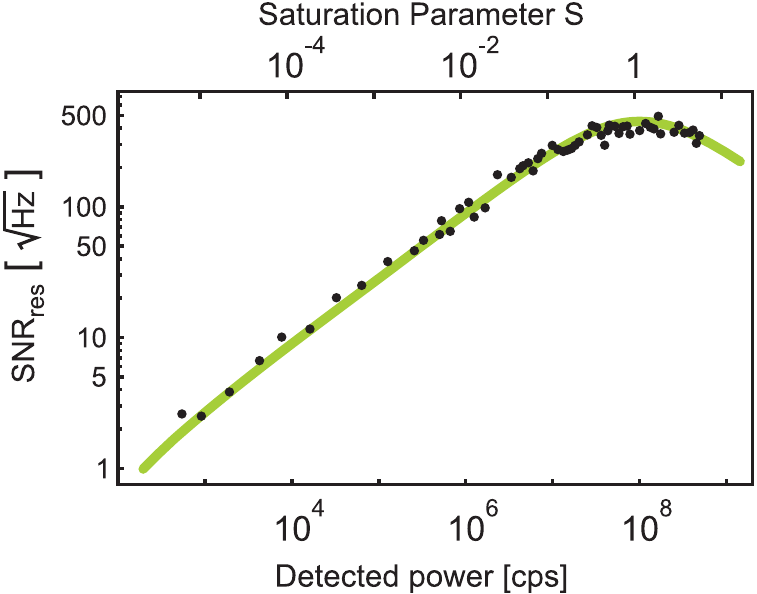}}
\caption{The signal-to-noise ratio of a transmission measurement at
the laser frequency as a function of the the laser power on the
detector and of the saturation parameter.
\label{intensity-dependence}}
\end{figure}

Considering that the solid angle $\Omega$ collects all the incident
laser light, a simple energy balance argument implies that $P_{\rm
ext}$ in Eq.~(\ref{mainformula}) must correspond to the total power
$P^{4\pi}_{\rm m}$ emitted by the molecule. Now we insert a spectral
filter to select only the part of the transmitted light that is
resonant with the laser light. Denoting the size of the dip in the
power that is detected in this case by $P^{\rm res}_{\rm dip}$,
Eq.~(\ref{mainformula}) and Eqs.~(\ref{bunch}) yield,
\begin{equation}
P^{\rm res}_{\rm dip}=P_{\rm ext}- P^{\rm res}_{\rm m}
=P^{4\pi}_{\rm m}-P^{\rm res}_{\rm m} = (1-\alpha \zeta)\frac {
\Gamma_1 } 2  \frac { S} {1+S}~.
\end{equation}
The noise on a resonant extinction measurement is composed of the
shot noise $\sqrt{P_{\rm las}}$ of the laser power, the laser
intensity fluctuations $\kappa
 P_{\rm las}$ where $\kappa$ is a proportionality constant, and $\sqrt{P_{\rm drk}}$. Since these contributions are
statistically independent, the total noise can be written as $N_{\rm
res}=\sqrt{ \mu P_{\rm las} +(\mu \kappa P_{\rm las})^{2}+P_{\rm
drk}}$, where again, $\mu$ accounts for losses and the detection
efficiency. Assuming that intensity fluctuations have
 been mastered at a sufficient level and that
 $\mu P_{\rm las} \gg P_{\rm drk}$, one finds $N_{\rm res}\simeq\sqrt{ \mu P_{\rm las}}$.
Thus, the signal-to-noise ratio for an extinction measurement
becomes
\begin{eqnarray}
\label{SNRext}
 {\rm SNR}_{\rm res}  =  \frac{ \mu P^{\rm res}_{\rm dip } }{ N_{\rm res }  }
    \simeq   (1 -\zeta \alpha) \frac{ \Gamma_1} 2 \sqrt{ \frac{ \mu \alpha {\cal K}  }{\Gamma_2 }} \frac{ \sqrt S}{
    1+S}~.
\end{eqnarray}
Fig.~\ref{intensity-dependence} presents $\rm {SNR_{\rm res}}$ as a
function of the detected laser power $\mu P_{\rm las}$ and of the
saturation parameter $S$. In each case, $S$ was directly derived
from the power broadened linewidth of the fluorescence excitation
spectrum. Our system could perform at the shot-noise limit down to
the sub Hertz bandwidth over the whole power range presented here.
The green theoretical fit curve is obtained using
Eq.~(\ref{SNRext}). With the parameters that have been independently
determined for our setup ($\mu=0.2$, $\cal K$=0.5, $\alpha=0.2$,
$\Gamma_2=\Gamma_1/2, \Gamma_1/2\pi=17~\rm MHz$), an excellent
agreement with the measured data is achieved. The deterioration of
$\rm {SNR_{\rm res}}$ under very strong excitation is clearly
visible and stems from the fact that for a quantum emitter, $P^{\rm
res}_{\rm dip}$ saturates at high incident powers. The maximal
attainable SNR in a shot-noise limited resonant detection then
becomes $\rm {SNR_{\rm res}}=\sqrt{\Gamma_1^2 \alpha {\cal K \mu} /(
16 \Gamma_2)}$ and occurs at S=1 if we assume $\zeta \alpha\ll 1$.

\begin{figure}[htb]
\centerline{\includegraphics[width=8cm]{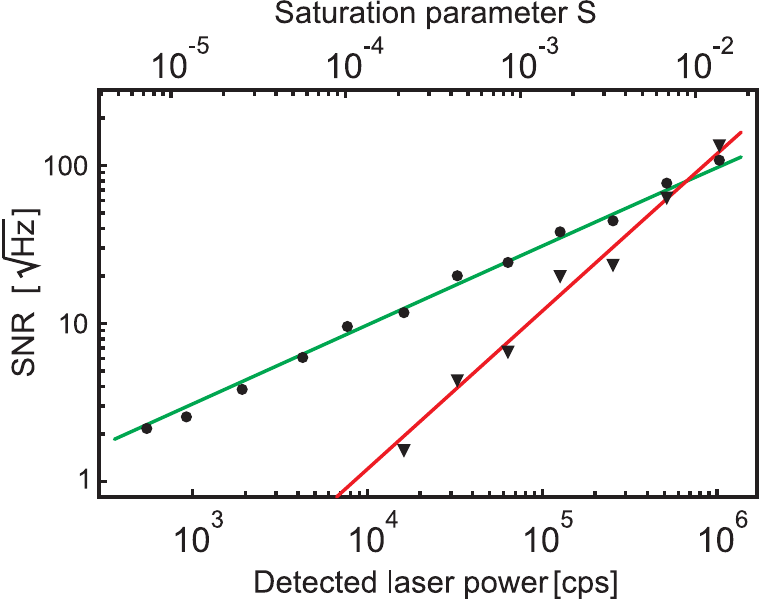}} \caption{The
signal-to-noise ratios of the resonant transmission (green) and
fluorescence (red) signals as a function of the excitation power in
the weak excitation regime. Symbols display the experimental data
and the lines denote theoretically expected behavior.
\label{comparison}}
\end{figure}

To compare the SNR of fluorescence and extinction measurements
directly, we have recorded spectra such as those shown in
Fig.~\ref{spectra} for low excitation powers corresponding to $S=6
\times 10^{-6}$ to $10^{-2}$ as shown in Fig.~\ref{comparison}. The
circles and the triangles display the SNR for the extinction and
fluorescence detections, respectively. To determine the experimental
SNR, we first fitted the spectra with Lorentzian functions. Then we
extracted the on-resonant signal and divided it by the off-resonant
rms noise. The data were recorded by adding 100 scans with 10~ms
integration time, corresponding to a total acquisition time of 1
second per frequency pixel. This procedure helped to correct for
possible laser drifts and spectral diffusion of the molecule. The
fitted green and red theoretical curves depend on $\sqrt{S}/(1+S)$
and $S/(1+S)$, respectively (see Eqs.~(\ref{SNRfluo}) and
(\ref{SNRext})) and show a very good agreement with the experimental
data. We conclude that in case of a weakly excited system, an
extinction measurement can be superior to fluorescence detection in
terms of SNR. We point out in passing that both $\rm SNR_{\rm red}$
and $\rm SNR_{\rm res}$ scale as the square root of the integration
time and thus, the comparison between the fluorescence and
extinction methods holds for fast and slow measurements alike.

We remark that in the existing literature, the root-mean-square
(rms) fluctuation of the signal itself has been often included as a
noise source in fluorescence but ignored in extinction
measurements~\cite{moerner06, plakhotnik:02}. In our comparison of
the two methods, we have consistently chosen to define the signal as
the response of the system of interest, namely a single emitter, and
the noise as all fluctuations stemming from other sources.
Therefore, we do not include the signal rms noise in our analysis of
SNR. This strategy is particularly convenient for the evaluation of
SNR from single spectra.

\begin{figure}[htb]
\centerline{\includegraphics[width=8cm]{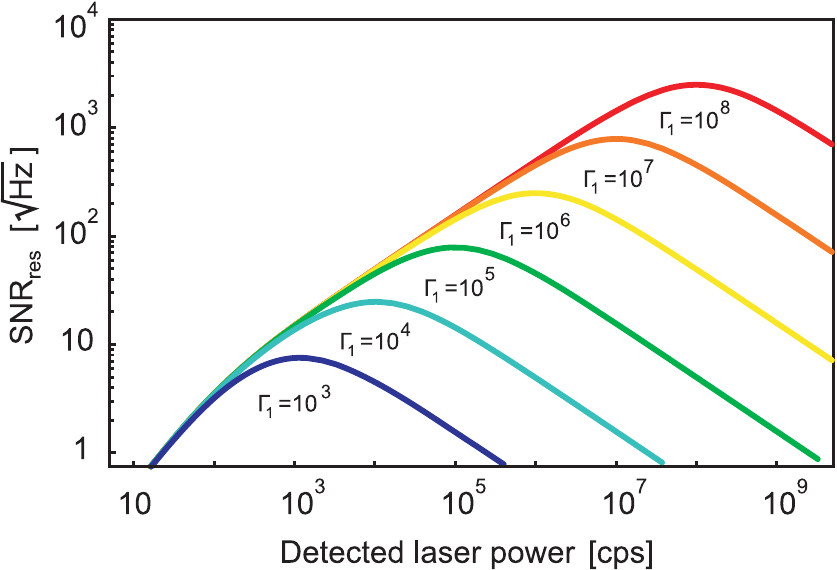}}
\caption{The SNR for a resonant transmission detection of emitters
with different radiative decay rates. Here we have assumed
$\alpha=\mu=1$, $P_{\rm drk}$=20 ~cps, and ${\cal K}=0.5$.
$\Gamma_1$ is given in units of $\rm rad/ sec$.
\label{linewidth-dependence}}
\end{figure}

An exciting question that arises is whether extinction detection
opens doors for studying weakly fluorescent nano-objects.
Conventional single molecule detection has been successful for
molecules that have fluorescence lifetimes of a few nanoseconds,
corresponding to $\Gamma_1/2\pi \sim 10-100~\rm MHz$. Such a high
photon flux provides a good SNR even considering realistic
collection plus detection efficiency of a few percent and $P_{\rm
drk}=100$ cps. However, for weakly emitting systems such as rare
earth ions with lifetimes of the order of milliseconds, $\rm
SNR_{red}$ becomes comparable or smaller than unity.
Fig.~\ref{linewidth-dependence} displays the expected $\rm SNR_{\rm
res}$ as a function of the detected laser power for various
radiative decay rates $\Gamma_1$. Here we have assumed a laboratory
value of ${\cal K}=0.5$, $\alpha=\mu=1$, $\Gamma_2=\Gamma_1/2$, and
$P_{\rm drk}=20~\rm cps$, but extension of the results to other
situations is straightforward by following Eq.~(\ref{SNRext}). These
plots indicate that single emitters with spontaneous emission times
as long as a millisecond should be detectable using extinction
spectroscopy even when realistic detection parameters (e.g. ${\cal
K}=0.5, \alpha=0.5, \mu=0.2)$ are considered. In addition, we
emphasize that extinction measurements have the great added value
that they provide direct access to the coherent interaction of the
incident light and the emitter.

Although the basic concepts discussed in this paper have been known
in signal processing and electrical engineering~\cite{Haus-new},
their direct experimental investigations at the single emitter level
have been made possible through advances in cryogenic
spectroscopy~\cite{Plakhotnik:01,Wrigge:08,Vamivakas:07}. Inspired
by this progress, very recently we have also succeeded in extinction
detection of a single solid-state quantum emitter at room
temperature~\cite{Kukura:08} despite the fact that the extinction
cross section is reduced by 5-6 orders of magnitude due to severe
broadening of the transition ($\Gamma_2 \gg \Gamma_1$). Another
interesting application of extinction or homodyne detection has been
demonstrated almost independently for imaging small metallic and
dielectric
nanoparticles~\cite{Mikhailovsky:03,Lindfors:04,Arbouet:04,Jacobsen:06,Ignatovich:06,Ewers:07}.
Conventional methods of nanoparticle detection such as
dark-field~\cite{Schultz:00} or total internal
reflection~\cite{Soennichsen:00} microscopy rely on the elimination
of the incident light from the detection path and the detection of
the power scattered by the particles. To this end, these techniques
are analogous to fluorescence excitation spectroscopy where
frequency spectra are replaced by spatial images, and spatial
filtering substitutes spectral filtering for the discrimination of
the incident laser light. However, in practice the two systems are
limited in different ways because in the case of spatial imaging,
the persistent source of "noise" is the light scattered from
residual optical roughness of the medium~\cite{Jacobsen:06}. The
equivalent of this problem usually does not arise in extinction
detection of emitters because they are typically embedded in
well-behaved matrices~\cite{Kulzer:04,Moerner:89} without any
optical transitions in the spectral region of interest.

In conclusion, we have explored the signal-to-noise ratio in the
spectroscopic detection of single emitters. We have provided
expressions for evaluating the performance of both fluorescence and
extinction measurements. In particular, we have demonstrated that
extinction measurements can be superior to fluorescence detection in
the weak excitation regime. Furthermore, we have shown that even
weakly fluorescent emitters should be detectable using coherent
extinction spectroscopy. This prospect is especially interesting for
the optical storage and read out of quantum information in new
systems such as rare earth ions.

\section*{Acknowledgments}\noindent
This work was supported by the Swiss National Foundation (SNF).

\end{document}